\documentclass[prl,twocolumn,floatfix,superscriptaddress]{revtex4}

 \usepackage{graphicx}
 \usepackage{bm}

 \newcommand{\be}{\begin{equation}}
 \newcommand{\ee}{\end{equation}}
 \newcommand{\ba}{\begin{eqnarray}}
 \newcommand{\ea}{\end{eqnarray}}
 \newcommand{\nn}{\nonumber \\}

\begin{document}

 \title{Role of Single Qubit Decoherence Time in Adiabatic Quantum Computation}
 \author{M.~H.~S.~Amin}
 \affiliation{D-Wave Systems Inc., 100-4401 Still Creek Drive, Burnaby, B.C., V5C 6G9, Canada}
 \author{C.~J.~S.~Truncik}
 \affiliation{D-Wave Systems Inc., 100-4401 Still Creek Drive, Burnaby, B.C., V5C 6G9, Canada}
 \author{D.~V.~Averin}
 \affiliation{Department of Physics and Astronomy, SUNY Stony Brook, Stony Brook NY 11794, USA}

 \begin{abstract}

 We have studied numerically the evolution of an adiabatic quantum
 computer in the presence of a Markovian ohmic environment by considering
 Ising spin glass systems with up to 20 qubits independently coupled
 to this environment via two conjugate degrees of freedom. The required
 computation time is demonstrated to be of the same order as that for an
 isolated system and is not limited by the single-qubit decoherence time
 $T_2^*$, even when the minimum gap is much smaller than the temperature
 and decoherence-induced level broadening. For small minimum gap,
 the system can be described by an effective two-state model
 coupled only longitudinally to environment.

 \end{abstract}
 \maketitle

Adiabatic quantum computation \cite{Farhi} (AQC) is an attractive
model of quantum computation (QC). It eliminates the need for
precise timing of the qubit transformations required in the
gate-model computation scheme, and also is expected to possess some
degree of fault tolerance afforded by the energy gap separating the
ground from excited states of the qubit Hamiltonian. AQC approach is
particularly appealing in the context of superconducting qubits
which in principle have the required flexibility for implementation
of complicated interactions. In the AQC, a system starts from a
readily accessible ground state of some initial Hamiltonian $H_i$
and slowly evolves into the ground state of the final Hamiltonian
$H_f$ which encodes solution to the problem of interest:
 \be
 H_S(t) = [1-s(t)] H_i + s(t) H_f, \label{HS}
 \ee
 where $s(t)\in [0,1]$ is a monotonic function of time $t$. Here, we
 only consider a linear time sweep $s(t)=t/t_f$, where
 $t_f$ is the total evolution time. Transitions out of the ground
 state can be caused by the Landau-Zener processes \cite{LZ} at the
 anticrossing ($s{=}s^*$), where the gap $g$ between the ground state
 $|0\rangle$ and first excited state $|1\rangle$ goes through a minimum:
 $g_m\equiv g(s^*)$. The probability of being in the ground state at
 the end of the adiabatic evolution is approximately ($\hbar=k_B=1$)
 \be
 P_{0f}=1-e^{- t_f/t_a}, \quad
 t_a \equiv {4 \over \pi g_m^2}\left|\langle 1| {dH_S \over ds}|
 0\rangle\right|_{s=s^*}.
 \label{acon}
 \ee
 To ensure large $P_{0f}$, one needs $t_f\gtrsim t_a$. The
 computation time is hence determined by $t_a$ and thus by $g_m$.

 In the gate-model QC, there is no direct correspondence between the
 wavefunction and the instantaneous system Hamiltonian. The
 Hamiltonian is only applied at the time of gate operations and
 usually involves only a few qubits. The wavefunction, therefore,
 is strongly affected by the environment and is irreversibly altered
 after the decoherence time, which is typically smaller than the
 single-qubit dephasing time $T_2^*$. This means that $T_2^*$ imposes
 an upper limit on the total computation time, unless some
 quantum error correction scheme (which requires significant resources)
 is utilized. This is not true for AQC, as the wavefunction is always
 very close to the instantaneous ground state of the system
 Hamiltonian and is consequently more stable against the decoherence.
 Qualitatively, one expects decoherence to drive the system's reduced
 density matrix towards being diagonal in the energy basis, which is
 not harmful for AQC but is detrimental for the gate-model QC. Such
 robustness has been demonstrated in previous studies
 \cite{Lidar,Childs,Roland2,Tiersch,Ashhab,Amin1,Amin2,Amin3}. However,
 those studies have either used a two-state model to describe
 the behavior of a multi-level system at the anticrossing, or
 assumed noise models that are not motivated by physical
 implementations. In this paper, we numerically study
 quantum evolution of a multi-qubit system directly without the
 two-state approximation, assuming a quite general and realistic
 coupling to environment.

 We consider a very general Hamiltonian $H(t) = H_S(t) + H_B + H_{\rm
 int}$, which includes system, bath, and interaction between them,
 respectively. The dynamics of the total (system plus environment)
 density matrix is governed by the Liouville equation \cite{Blum}:
 $\dot \rho (t) = -i [H(t), \rho(t)]$. The reduced density matrix for
 the system is obtained by partially tracing over the environmental
 degrees of freedom: $\rho_S=$Tr$_B[\rho]$.
 Let $|n(t)\rangle$ denote the instantaneous eigenstates of the
 system Hamiltonian: $H_S(t) |n(t)\rangle = E_n(t) |n(t)\rangle$. In
 this basis, we define $\rho_{nm}(t) = \langle
 n(t)|\rho_S(t)|m(t)\rangle$. Taking the time derivative, we obtain
 (dropping explicit time dependences)
 \ba
 \dot \rho_{nm} = \langle n|\dot \rho_S |m \rangle +
 \langle \dot n|\rho_S|m \rangle + \langle n|\rho_S|\dot m
 \rangle .  \label{ad1}
 \ea

 We begin by focusing on the first term in (\ref{ad1}) which is
 responsible for the decay processes. We treat it
 quasi-statically, assuming that the evolution of the
 Hamiltonian is much slower than the environmentally-induced decay
 rates, so that the eigenstates can be taken as time independent.
 We also assume that the effect of the system on the environment
 is so small that the bath maintains its equilibrium distribution
 $\rho_B$ at all times. Moreover, the bath is taken to have
 correlation time $\tau_B$ shorter than all the decay times of the system
 so that one can apply Markovian approximation. Then, using the standard
 Bloch-Redfield formalism, one can show that \cite{Blum,Weiss}
 \ba
 \langle n| \dot \rho_{S} |m \rangle &=&  -i\omega_{nm}
 \rho_{nm} + e^{-i \omega_{nm} t}\langle n| \dot \rho_{SI} |m
 \rangle \nn
 &=& - i\omega_{nm}\ \rho_{nm}
 - \sum_{k,l} R_{nmkl} \ \rho_{kl}, \label{BReq}
 \ea
 where $\omega_{nm}=E_n-E_m$, and
 \ba
 && R_{nmkl} = \delta_{lm} \Gamma^{(+)}_{nrrk} +
 \delta_{nk} \Gamma^{(-)}_{lrrm}
 -\Gamma^{(+)}_{lmnk} - \Gamma^{(-)}_{lmnk}, \quad
 \label{R} \\
 &&\Gamma^{(+)}_{lmnk} = \int_0^\infty dt \ e^{-i\omega_{nk}t}
 \langle \tilde{H}_{I,lm}(t)\tilde{H}_{I,nk}(0) \rangle, \nn
 &&\Gamma^{(-)}_{lmnk} = \int_0^\infty dt \ e^{-i\omega_{lm}t}
 \langle \tilde{H}_{I,lm}(0)\tilde{H}_{I,nk}(t) \rangle,
 \label{Gammapm} \\
 &&\tilde{H}_{I,nm} (t) = \langle n| e^{iH_Bt} H_{\rm int} (t)
 e^{-iH_Bt}|m\rangle. \nonumber
 \ea
 Here, $\langle ... \rangle \equiv \text{Tr}_B[\rho_B ...]$, and
 summation over repeated indices is implied in (\ref{R}).

 Substituting (\ref{BReq}) into (\ref{ad1}), we obtain
 \ba
 \dot \rho_{nm} = -i\omega_{nm} \rho_{nm} -\sum_{k,l}
 \left( R_{nmkl} -  M_{nmkl} \right) \rho_{kl}, \label{GBR}
 \ea
 where $M_{nmkl} = \delta_{nk} \langle l| \dot m \rangle +
 \delta_{ml} \langle \dot n|k \rangle$. The tensors $M_{nmkl}$ and
 $R_{nmkl}$ are responsible for non-adiabatic and thermal
 transitions, respectively. For a time-independent Hamiltonian,
 $M_{nmkl}=0$, and (\ref{GBR}) becomes the Bloch-Redfield equations
 \cite{Blum,Weiss}. The derivatives like $|\dot n \rangle$ can be calculated
 numerically. It is important to ensure that the equation stays
 trace-preserving, which requires ${\rm Re}\sum_{n,m} \langle n|\dot m
 \rangle = 0$. This condition is {\em exactly} satisfied (even with
 the truncation discussed below), if we write $\langle n(t) |\dot
 m(t)\rangle = {1 \over 4\delta t} \{\langle n(t{+}\delta t)|+
 \langle n(t{-}\delta t)|\} \{|m(t{+}\delta t) \rangle -|m(t{-}\delta
 t) \rangle \}$.
 %
 %
 %\ba
 % \langle n(t) |\dot m(t)\rangle = {1 \over 4\delta t} \{\langle n(t+\delta t)|
 % + \langle n(t-\delta t)|\} \nn
 % \times \{|m(t+\delta t) \rangle -|m(t-\delta t) \rangle \}.
 %\ea
 %

 To introduce coupling to environment, we consider a quite general
 interaction Hamiltonian
 \be
 H_{\rm int} = - \sum_{i=1}^n \left(Q_x^{(i)} \sigma_x^{(i)}
 + Q_z^{(i)} \sigma_z^{(i)}\right), \label{coupl}
 \ee
 where $\sigma_{x,z}^{(i)}$ are the Pauli matrices of the $i$-th qubit,
 and $Q_\alpha^{(i)}$ are the heat-bath operators. Using
 (\ref{Gammapm}), and assuming uncorrelated heat baths, we find
 \ba
 \Gamma^{(+)}_{lmnk} &=&  {1\over 2} \sum_{i,\alpha} S_\alpha^{(i)} (-\omega_{nk})
 \sigma_{\alpha,lm}^{(i)} \sigma_{\alpha,nk}^{(i)}, \nn
 \Gamma^{(-)}_{lmnk} &=&  {1\over 2}  \sum_{i,\alpha} S_\alpha^{(i)} (\omega_{lm})
 \sigma_{\alpha,lm}^{(i)} \sigma_{\alpha,nk}^{(i)},
 \ea
 where
 %
 %\ba
 $ S_\alpha^{(i)}(\omega) = \int_{-\infty}^\infty dt \ e^{i\omega t}
 \langle Q_\alpha^{(i)} (t) Q_\alpha^{(i)} (0) \rangle$
 %\ea
 %
 are the bath spectral densities and  $\sigma_{\alpha,lm}^{(i)} =
 \langle l| \sigma_{\alpha}^{(i)} |m\rangle$. Here, we have neglected
 the imaginary parts of $\Gamma^{(\pm)}_{lmnk}$, as they only produce
 small shifts of energies which in principle can be accounted for by
 proper renormalization.

 To model the spectral densities, we assume ohmic bosonic heat baths
 in thermal equilibrium \cite{Leggett}: $S_\alpha^{(i)}(\omega)
 = \eta_\alpha^{(i)}\omega e^{-|\omega|/ \omega_c}/(1-e^{-\omega/T})$.
%    %
%    \be
%    S_\alpha^{(i)}(\omega) = \eta_\alpha^{(i)}{ \omega
%    e^{-\omega/ \omega_c} \over 1-e^{-\omega/T}}.
%    \ee
%    %
 The dimensionless coefficients $\eta_\alpha^{(i)}$ describe the
 strength of coupling between the qubits and environment, and
 $\omega_c$ is a cutoff frequency which we assume to be larger than
 all relevant energy scales. The Markovian approximation is valid as
 long as $\tau_B \sim 1/\omega_c$ is shorter than all the decay times
 and the characteristic time variation of the Hamiltonian.

\begin{figure}[t]
 \includegraphics[width=7.5cm]{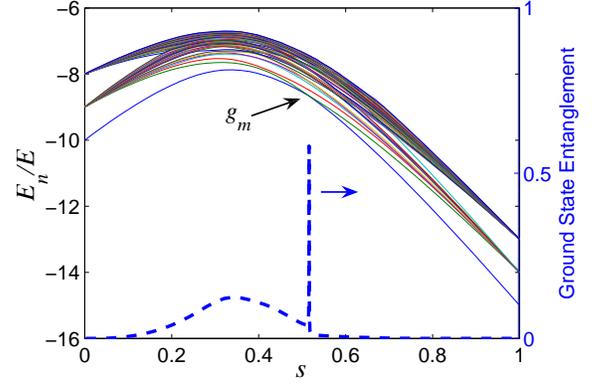}
 \caption{(Color online). Energy spectrum for a 20 qubit instance with a very small minimum
 gap: $g_m/E\approx 5 \times 10^{-4}$. Only the first 50 energy levels
 of the total $\sim 10^6$ are shown. The dashed line in the bottom represents
 the ground state entanglement calculated using Meyer and Wallach
 entanglement measure (\ref{MW}).}\label{fig1}
 \end{figure}

 We now use the above model to study the evolution of a multi-qubit
 Ising system with the initial and final Hamiltonians given by
 \ba
 &&{H_i \over E} = -{1 \over 2}\sum_i \Delta_i \sigma_x^{(i)}, \\
 &&{H_f \over E} = -{1 \over 2}\sum_i h_i \sigma_z^{(i)}
 + {1 \over 2}\sum_{i>j} J_{ij} \sigma_z^{(i)}\sigma_z^{(j)},
 \ea
 where $\Delta_i$, $h_i$, and $J_{ij}$ are dimensionless
 parameters and $E$ is an energy scale. We consider square lattice
 configurations with nearest and next-nearest neighbor couplings
 between the qubits. The choice of only 2-qubit short-range
 interactions is motivated by the feasibility of experimental
 implementation. We generate spin glass instances involving 6, 9,
 12, 16, and 20 qubits by randomly choosing $h_i$ and $J_{ij}$ from
 $\{-1, 0,1\}$ and identifying small gap instances with
 non-degenerate final ground state (see, e.g., Fig.~\ref{fig1}). Such
 instances are very rare and represent difficult problems; a
 degenerate ground state (multiple solutions) ensures higher
 probability of finding one of the solutions. We also choose
 $\Delta_i=1$ for all $i$.

 Figure \ref{fig1} shows the energy spectrum for a 20 qubit
 instance with a very small $g_m$. The first two energy levels anticross
 near the middle of the evolution. We should mention that this is
 not a typical 20 qubit spectrum for the problem we consider. Indeed
 for an average problem
 the minimum gap is much larger and instances with such a
 small gap are very rare. For such instances,
 the bottleneck of the adiabatic evolution is expected to be
 near the anticrossing. Moreover, since the gap is much smaller than
 the typical energy separation of the levels, two state
 approximation is expected to be sufficient to describe the
 evolution. This can be tested by comparing a fully numerical
 simulation without two state approximation with a two state model,
 which will be done at the end of this article.

 To study the evolution, we numerically integrate
 Eq.~(\ref{GBR}) starting from $\rho_S(0)=|0\rangle \langle
 0|$. For large number of qubits, the computation becomes extremely
 time consuming because of the large number of matrix elements in
 $\rho_S$. However, since $\rho_S$ is written in the energy basis at
 all times, it is possible to significantly simplify the computation
 by truncating $\rho_S$ to only the lowest few energy levels occupied
 in the course of evolution. To ensure small error, we
 increase the number of levels kept in the calculation until the
 results do not change. Typically maximum 7-8 energy levels are
 sufficient for the type of evolution and instances we consider here.
 For very slow evolutions, even two states suffice to achieve
 acceptable accuracy. The numerical integration time can also be
 significantly reduced by refining the integration steps based on the
 gap size.

 \begin{figure}[t]
 \includegraphics[width=7.5cm]{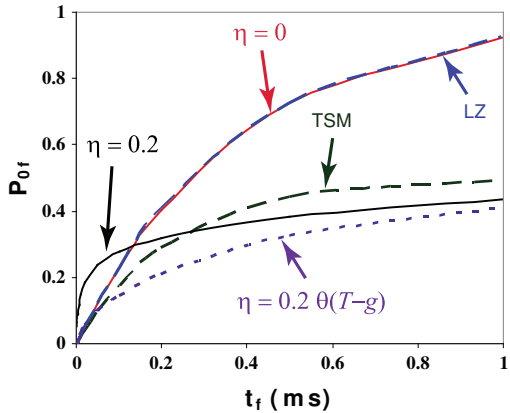}
 \caption{(Color online). Probability of success $P_{0f}$ as a function of $t_f$ for
 the 20 qubit instance of Fig.~\ref{fig1}. The solid lines are
 calculated  with ($\eta=0.2$) and without ($\eta=0$) coupling to the
 environment. Other parameters are $E=10$ GHz, $T=25$ mK. The case
 without coupling to the environment is compares with the pure
 Landau-Zener behavior using (\ref{acon}) (see dashed blue
 line). The excellent agreement confirms that for the coherent
 evolution, the two state model is sufficient to calculate the
 probability. The (black) dashed line, marked by TSM, is obtained
 using analytical formula (\ref{analy}) obtained using a two state model.
 It asymptotically shows the same behavior as the numerical result
 (solid line marked by $\eta=0.2$).
 The dotted line is numerical calculation with $\eta =
 0.2 \ \theta(T-g)$, which eliminates relaxation after the anticrossing
 (see the text for description).}\label{fig2}
 \end{figure}

 Figure \ref{fig2} shows the probability of staying in the ground
 state at the end of the evolution as a function of $t_f$ for the 20
 qubit instance depicted in Fig.~\ref{fig1}. For simplicity, we have
 chosen the same coupling to the environment for all qubits:
 $\eta_\alpha^{(i)}=\eta$. Notice that with the chosen parameters,
 the minimum energy gap is much smaller than temperature ($g_m/T \approx 10^{-2}$),
 hence thermal excitation at the anticrossing
 is expected. For a closed system ($\eta=0$) the numerics agree very
 well with Eq.~(\ref{acon}). For an open system ($\eta=0.2$), however, the
 probability is enhanced (compared to the closed system) at small $t_f$,
 while it is suppressed for large $t_f$, asymptotically approaching its
 equilibrium value. Therefore, it is more efficient
 to run the system for a shorter time and repeat the process than to
 wait for a long time to attain significant probability. An important
 point is that the time scale for the probabilities to reach some
 non-vanishing value is almost the same ($\sim$ 1 ms) for all curves.
 This has been a generic property for all instances that we have studied
 regardless of the size of the gap.

 The thermally assisted behavior
 in short $t_f$ regime is the result of large relaxation after
 the anticrossing region and is not expected to enhance the scaling
 of the computation \cite{Amin1}. To confirm this, we have repeated
 the numerical calculations, but now allowing transitions only in the
 thermal mixing region by choosing $\eta=0.2\ \theta(T-g)$. This type of
 coupling coefficient only allows thermalization in a region with
 $g<T$ and therefore eliminates
 the relaxation back to the ground state after the anticrossing.
 The result (dotted line in Fig.~\ref{fig2}) shows no initial
 enhancement compared to the closed system, confirming the above statement.

 We now compare the numerically calculated computation time with
 the single-qubit decoherence times. If the qubits are uncoupled
 ($J_{ij}{=}0$), the single-qubit decoherence rates, in weak
 coupling limit, are given by $1/T_2^* \sim S_\alpha^{(i)}(0)\sim \eta T$.
 %$\Gamma_2^{(i)}=\gamma_i/2+(\sin\theta_i)^2
 %\tilde{S}_x^{(i)}(0)+(\cos\theta_i)^2 \tilde{S}_z^{(i)}(0)$, where
 %$\gamma_i= (\cos\theta_i)^2
 %\tilde{S}_x^{(i)}(\Omega)+(\sin\theta_i)^2
 %\tilde{S}_z^{(i)}(\Omega)$ is the relaxation rate, $\tan
 %\theta_i=(1-s)\Delta_i/sh_i$,
 %$\tilde{S}_\alpha^{(i)}(\omega)=S_\alpha^{(i)}(\omega) +
 %S_\alpha^{(i)}(-\omega)$ is the symmetrized spectral density, and
 %$\Omega=\sqrt{(1-s)^2\Delta_i^2+s^2h_i^2}E$ is the energy gap.
 For the parameters of Fig.~\ref{fig2}, $T_2^*{\sim} 10$~ns, which
 is typical for solid-state qubits. This decoherence time is five
 orders of magnitude smaller than the computation time (${\sim} 1$
 ms) for the problem of Fig.~\ref{fig2}. Therefore, unlike the gate
 model QC, {\em in AQC, the computation time is not limited by the
 single-qubit decoherence time}.

 It should be noted that the qubits
 will go through an entangled state during the evolution.
 To demonstrate that, we have displayed in Fig.~\ref{fig1}
 the ground state entanglement (dashed line) calculated using the measure
 originally proposed by Meyer and Wallach \cite{Meyer}:
 \be
 Q(|\psi\rangle)={1\over n}\sum_{k=1}^n 2(1-\text{Tr}[\rho_k^2])\,
 ,   \label{MW}
 \ee
 where $\rho_k\equiv\text{Tr}_{j\neq k}|\psi\rangle \langle \psi|$ is
 obtained by partially tracing over all qubits except the $k$-th one.
 The ground-state entanglement
becomes nonzero in the first half of the evolution with a very sharp
peak at the anticrossing. The measure (\ref{MW}), however, does not
describe
 quantitatively the actual (mixed-state) entanglement at that point,
 since it does not account for thermal mixing. Unfortunately, no
 practical mixed-state entanglement measure exists for more than
 two qubits \cite{Wootters}. The entanglement will not
 be destroyed by the environment (except maybe at the anticrossing)
 as long as the system dominantly populates the ground state.
 Therefore, despite the fact that the evolution time is far beyond
 the qubits' dephasing time, the system still preserves its quantum
 mechanical behavior throughout the evolution. Once
 again, this is in contrast to what is expected in the gate-model QC.
 Qualitative demonstration of the
 non-vanishing ground state entanglement during the evolution is
 also important as it shows that the evolution cannot be
 described efficiently in only classical terms.

\begin{figure}[t]
 \includegraphics[width=7cm]{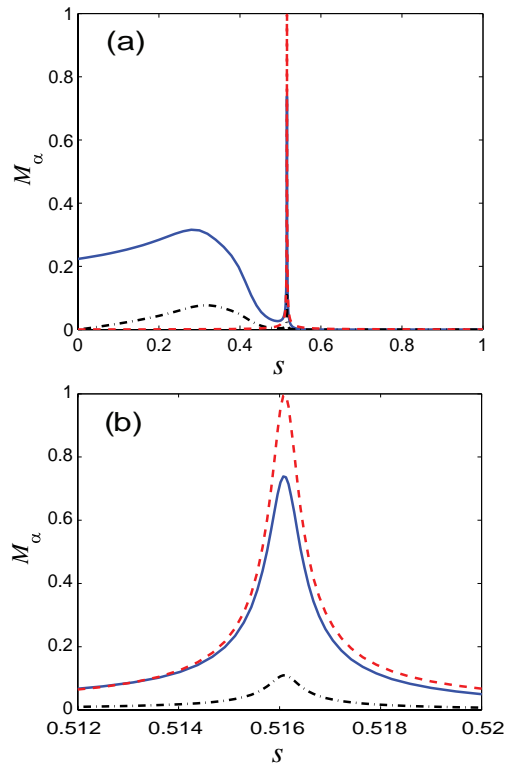}
 \caption{(Color online). (a) The rms value of the matrix elements of the Pauli
 matrices between the first two states, as define in (\ref{M}).
 Solid (blue) line
 is $M_z$, dashed-dotted (black) line is $M_x$. We have also plotted
 the matrix element ($M_z^{\rm TSM}$) of $\tau_z$ between
 the two energy levels of a two
 state model described by Hamiltonian (\ref{HTSM}).
 (b) The
 same curves zoomed near the anticrossing show qualitative
 agreement (except for a prefactor) between all three curves. }\label{fig3}
 \end{figure}

 The numerical method presented here is valid only for a Markovian
 environment. Most environments, however, especially in
 superconducting systems, are not Markovian and have a significant
 amount of low frequency noise. In Ref.~\cite{Amin3},
 we have used the two-state model (TSM) approximation
 to study the effect of a non-Markovian environment on AQC. For the
 rest of this paper, we focus on showing that the TSM is adequate for
 the description of the AQC performance in the small-gap regime.
 For the environment to be able to cause transitions out of
 the ground state, the interaction Hamiltonian should have nonzero
 matrix elements between the ground state and the target state. We
 introduce
 \be
 M_\alpha = \left({1\over
 n}\sum_{i}|\sigma_{\alpha,10}^{(i)}|^2\right)^{1/2} \label{M}
 \ee
 which give the rms values of the matrix elements of Pauli matrices
 $\sigma_\alpha^{(i)}$ between the lowest two states. They represent
 some average behavior of the corresponding matrix elements
 and are closely related to relaxation between the two levels.
 Especially, if all the qubits have the same coupling $\eta$ to
 the environment, the relaxation rate between the two states is
 given by
 \be
 \gamma = n(M^2_x+M^2_z)\tilde{S}(\omega_{10}), \label{gamma}
 \ee
 where $\tilde{S}(\omega)$ is the symmetrized spectral density of the
 (uncorrelated) baths. Figure \ref{fig3}(a) displays $M_x$ and $M_z$
 as a function of $s$ for the 20-qubit system of Fig.~\ref{fig1}. Except for
 the initial region, they both show the same behavior: a sharp peak
 at the anticrossing, with a width proportional to $g_m$, followed by
 a vanishingly small value. For small $T$, the excitation from the
 ground state will be suppressed everywhere except near the
 anticrossing, where $g<T$.

 The transitions at the anticrossing can be described by an effective
 two-state Hamiltonian:
 \ba
 &&H_S^{\rm TSM} = -{1\over 2}(g_m \tau_x + \epsilon \tau_z), \nn
 &&H_{\rm int}^{\rm TSM} = -(Q_x^{\rm TSM} \tau_x + Q_z^{\rm TSM}
 \tau_z), \label{HTSM}
 \ea
 where $\tau_\alpha$ are the Pauli matrices in the two-state
 subspace, $\epsilon = 2\tilde{E}(s-s^*)$, with $s^*$ being the
 position of the anticrossing and $\tilde{E}$ an energy scale (in
 our case, close to $E$) characterizing the anticrossing. We
 introduce the matrix elements
 \ba
 &&M^{\rm TSM}_x = |\langle 0|\tau_x|1\rangle| = |\cos \theta|, \nn
 &&M^{\rm TSM}_z = |\langle 0|\tau_z|1\rangle| = |\sin \theta|,
 \label{MTSM}
 \ea
 which are
 responsible for the transitions between the two levels, where $\tan
 \theta = g_m/\epsilon$. Except for an
 overall factor, $M^{\rm TSM}_z$ has the same shape as both $M_{x,z}$
 (see the inset of Fig.~\ref{fig3}), while $M^{\rm TSM}_x$ is completely
 different; it has a sharp dip at $\epsilon=0$ where it vanishes.
 Thus, in order for the effective
 two-state model to give the same result as the full system, it
 needs to couple to the environment only via $\tau_z$ (i.e., $Q_x^{\rm
 TSM}=0$). In other words, a generic single-qubit coupling to the environment
 reduces to only longitudinal coupling in the effective TSM. Besides
 the numerical agreement, the fact that $Q_x^{\rm TSM}=0$ in the two
 state model has a deep physical meaning. As is clear in
 (\ref{MTSM}), $M_x^{\rm TSM} \approx 1$ everywhere except for a very small
 region near the anticrossing. In the presence of a non-vanishing $Q_x^{\rm
 TSM}$, the relaxation rate will be very large almost everywhere and
 therefore the system should relax to the ground state in constant
 time even when $s=1$. This is obviously unphysical and especially
 not expected for spin glasses for which the relaxation time to the
 ground state is extremely long. In the appendix of
 Ref.~\cite{Amin2}, the effective Hamiltonian (\ref{HTSM}) is
 systematically derived for the case of adiabatic Grover search
 problem \cite{Roland}. For that problem, one finds $Q_x^{\rm TSM}=O(1/\sqrt{N})$, but
 $Q_z^{\rm TSM}=O(1)$, and therefore in large $N$ limit the former
 vanishes in agreement with our physical expectation. Notice that
 in this case, the relaxation rate due to coupling of the bath to
 $\tau_x$ is $\gamma \propto (Q_x^{\rm TSM})^2 = O(1/N)$,
 therefore solving the problem merely based on relaxation leads
 to a computation time $t_f=O(N)$, which is the complexity of
 classical computation.

 From the TSM Hamiltonian (\ref{HTSM}) with this type of
 longitudinal coupling, the success probability in the
 large-$T$ limit is \cite{Amin1,Amin3}
 \be
 P_{0f}^{\rm TSM}={1\over 2}(1-e^{-2t_f/t_a}). \label{analy}
 \ee
 This formula is also plotted in Fig.~\ref{fig2}. The qualitative
 asymptotic agreement with other numerical curves in the figure indicates
 that most of the
 transition occur in the small-gap region ($g \ll T$) where the TSM is
 adequate for their description.

 To summarize, by studying spin glass instances of up to 20
 qubits (only one is illustrated), we have explicitly demonstrated
 that the computation time in AQC can be much longer than the
 single-qubit decoherence time $T_2^*$. In the case of small
 minimum gap (i.e., hard instances), effective two-state model
 with only the longitudinal coupling to environment
 describes transitions at the anticrossing. The numerical results
 also show that the computation time scale is unaffected by ohmic
 environment. This conclusion can not be understood directly as
 suppression of transitions to the excited states by the energy gap,
 since in the chosen instances it was much smaller at the
 anticrossing than the temperature and decoherence strength. Rather,
 it arises from the balance of transitions between the two lowest
 levels. It should be emphasized that these results were obtained under
 the assumption of weak coupling to the environment, for which the
 discrete energy structure of $H_S$ is mostly preserved. In the case
 of strong coupling, the interaction Hamiltonian would dominate, and
 both the method and the conclusions of this work would not hold.

 The authors are grateful to A.J.~Berkley, P.~Bunyk, V.~Choi,
 R.~Harris, J.~Johansson, M.W.~Johnson, T.M. Lanting,
 S.~Lloyd, and G.~Rose for useful
 discussions.


\begin{thebibliography}{99}

 \bibitem{Farhi} E. Farhi, J. Goldstone, S. Gutmann, J. Lapan, A.
 Lundgren, and D. Preda, Science {\bf 292}, 472 (2001).

 \bibitem{LZ} L.D.~Landau, Phys.\ Z.\ Sowjetunion {\bf 2}, 46 (1932);
 C.~Zener, Proc.\ R.\ Soc. A {\bf 137}, 696 (1932).

 \bibitem{Lidar} M.S. Sarandy and D. Lidar, Phys.\ Rev. A {\bf 71},
 012331 (2005); Phys. Rev. Lett. {\bf 95}, 250503 (2005).

 \bibitem{Childs} A.M. Childs, E. Farhi, and J. Preskill, Phys.\ Rev. A {\bf 65},
 012322 (2001).

 \bibitem{Roland2} J. Roland and N.J. Cerf, Phys.\ Rev. A {\bf 71}, 032330
 (2005).

 \bibitem{Tiersch}  M.~Tiersch and R.~Sch\"utzhold, Phys.\ Rev. A {\bf 75},
 062313 (2007).

 \bibitem{Ashhab}  S. Ashhab, J.R. Johansson, and F. Nori,
 Phys. Rev. A {\bf 74}, 052330 (2006).

 \bibitem{Amin1} M.H.S.~Amin, P.J.~Love, and C.J.S.~Truncik,
 Phys.\ Rev.\ Lett. {\bf 100}, 060503 (2008).

 \bibitem{Amin2} A.T.S.~Wan, M.H.S.~Amin, and S.X.~Wang,
 Int. J. Quant. Inf. {\bf 7}, 725 (2009).

 \bibitem{Amin3} M.H.S.~Amin, D.V.~Averin, and J.A.~Nesteroff,
 Phys. Rev. A {\bf 79}, 022107 (2009).

 %\bibitem{Argyres} P.N. Argyres and P.L. Kelley, Phys. Rev. {\bf 134}, A98 (1964).

 \bibitem{Blum} K. Blum, ``Density Matrix Theory and Applications'',
 Plenum Pub. Corp., New York, 1st edition (1981).

 \bibitem{Weiss} U. Weiss, ``Quantum Dissipative Systems'', World Scientific,
 Singapore, 2nd edition (1999).

 \bibitem{Leggett} A.J. Leggett {\em et al}., Rev.\ Mod.\ Phys. {\bf 59}, 1
 (1987).

 \bibitem{Meyer} D.A.~Meyer and N.R.~Wallach, J.\ Math.\ Phys. {\bf
 43}, 4273 (2002).

 %\bibitem{Brennen} G.K. Brennen, Quant. Inf. Comput. {\bf 3}, 616
 %(2003).

 \bibitem{Wootters} W.K.~Wootters, Phys.\ Rev.\ Lett. {\bf 80}, 2245 (1998).

 \bibitem{Roland} J. Roland and N.J. Cerf, Phys. Rev. A {\bf 65}, 042308
 (2002).


 \end{thebibliography}
 \end{document}